# Equivalent Circuit Programming for Estimating the State of a Power System

Marko Jereminov, Martin R. Wagner
Carnegie Mellon University
Department of ECE
Pittsburgh, USA

Aleksandar Jovicic, Gabriela Hug
ETH Zurich
Power System Laboratories
Zurich, Switzerland

Larry Pileggi
Carnegie Mellon University
Department of ECE
Pittsburgh, USA

*Abstract* — An Equivalent Circuit Programming (ECP) approach that expresses the optimality conditions of an optimization problem in terms of an equivalent circuit model and uses circuit simulation techniques to solve for an optimal solution, is applied to the state estimation problem for power systems. The benefits of using an equivalent circuit formulation for incorporating both Phasor Measurement Units (PMU) and Remote Terminal Units (RTU), as well as for reducing the nonlinearities of the state estimation problem was previously demonstrated. In this paper we further exploit the circuit nature of the state estimation problem to formulate not only the model but also the optimality conditions as an ECP problem. The efficiency and accuracy of our approach are demonstrated by estimating the states of large-scale power grids (80k+ buses).

## I. Introduction

The reliable operation and future planning of the modern transmission level power system is highly dependent on efficient and accurate analysis of its steady-state behavior. Importantly, along with various sources of uncertainty [1], the increase of distributed generation, as well as load variations and inexact network topology information, there is a significant amount of inherent inaccuracy in the modeling of power system operations. These uncertainties make the problem of estimating and analyzing the steady-state operating point of a power system increasingly challenging [1]. Therefore, in order to ensure reliable and efficient grid operations, it is of utmost importance to provide an accurate and efficient methodology for estimating its state that is compatible with the measurement data within the power grid.

The most commonly used formulation for power system state estimation (SE) was conceived several decades ago by Schweppe and Wildes [2]. The Weighted Least Square (WLS) algorithm was proposed based on inherently nonlinear power mismatch equations which are suited for Remote Terminal Unit (RTU) measurements composed of voltage magnitudes and active and reactive power flows. Recently, the state estimation area however has been undergoing significant changes due to increased deployment of Phasor Measurement Units (PMUs) that provide highly accurate current and voltage phasor measurements. When the system is fully observable by PMUs and the problem is formulated as a function of voltages and currents [3], the state estimation problem becomes linear. However, this scenario is unlikely to happen any time soon due to the cost of PMUs. Therefore, several hybrid formulations [4]-[7] have been proposed in an attempt to incorporate both PMU and RTU measurements within the state estimation framework. Most importantly, all of the existing single and multi-stage hybrid approaches represent the approximation and modification of the conventionally formulated problem. Therefore, efficient real-time state estimation that includes accurate power grid models and emerging grid technologies remains a challenging problem.

We have recently introduced the formulation for the steady-state analysis of power systems via an equivalent split-circuit for power flow [8]-[13] and three-phase power flow problems. It was shown that the use of current and voltage state variables allows for a representation of the complete problem in terms of equivalent split-circuit models thereby enabling methods developed for circuit simulation of massive size circuits [14]-[15] to be adapted and applied for robust and efficient simulation of power grids [12]-[13]. Importantly, the current and voltage state variables are directly compatible with newly available grid measurement data from PMUs. This has recently led to the introduction of equivalent circuit representations for measurement devices such as the PMUs and RTUs to redefine the constraints of the power system State Estimation problem [16]. It was demonstrated that in addition to a significant decrease in the problem nonlinearities, the proposed state estimation problem formulation can simultaneously treat both PMU and RTU measurements within the same framework.

The optimization of power system steady-state behavior, namely, operation, state estimation, generated power dispatch, etc., is traditionally performed by defining the objective function that is to be minimized while satisfying the network, operational and/or stability constraints as traditionally formulated in the power flow problem. The defined problem is then generally implemented in one of the generalized nonlinear optimization toolboxes to obtain the optimal solution. However, we recently demonstrated [17]-[18] that the optimality conditions of power system optimization problems that are formulated in terms of equivalent circuit constraints will exhibit a unique characteristic: they represent the governing equations of a new equivalent circuit that consists of

This work was supported in part by the Defense Advanced Research Projects Agency (DARPA) under award no. FA8750-17-1-0059 for the RADICS program, and the National Science Foundation (NSF) under contract no. ECCS-1800812.

an original circuit and its adjoint circuit [17]. The operating point of such a circuit represents an optimal solution of the optimization problem and can be obtained as a solution to the circuit simulation problem. More broadly speaking, this formulation establishes a new class of optimization problems, namely Equivalent Circuit Programming (ECP) problems, for which constraints can be expressed in terms of equivalent circuit equations and state variables. Importantly, the ECP optimality conditions represent the governing equations of an equivalent circuit that is derived from the Tellegen's Theorem [15] and a generalization of adjoint network theory [19].

In this paper, we define the recently introduced power system SE formulation [16] as an ECP problem and show that it can be efficiently and accurately solved as a circuit simulation problem. Importantly, it can be demonstrated that the estimated states obtained from a solution of the ECP problem exactly match the ones from the commercial nonlinear optimization toolboxes. In addition, the circuit nature of ECP allows for a complete understanding of the nonlinearities introduced by the RTU circuit models, and enables the application of the recently developed power flow circuit simulation heuristics [12]-[13] to ensure robust simulation convergence and scalability.

We start with an overview of the equivalent split-circuit modeling of power system steady-state behavior with inclusion of the PMU and RTU measurement data. We then describe the Equivalent Circuit Programming by providing its generalized formulation as derived from Tellegen's Theorem (TT) [15] and adjoint network theory [19]. Furthermore, the ECP models of PMU and RTU measurement data are derived and hierarchically combined with the other power system elements to form an equivalent circuit whose operating point represents the estimated power system state in terms of current and voltage state variables. Lastly, the efficiency and scalability of the ECP formulation for SE is demonstrated on large-scale power grids, including the Eastern Interconnection tests cases.

## II. Split-Circuit formulation for modeling the power grid with PMU and RTU measurement devices

### A. Equvalent split-circuit modeling framework

Modeling the power system steady-state in terms of the traditional 'PQV' formulation [2]-[3] lacks direct compatibility with the measured data of currents and voltages [3] as it is based on power mismatch equations that are inherently nonlinear even though the underlying transmission network constraints are actually linear in nature (RLC circuit). In contrast, the network constraints within the equivalent circuit formulation are linear, defined in terms of current and voltage state variables, as they are directly derived from Kirchhoff's laws. The nonlinearities introduced by the commonly used generator and load models are translated to constraining the constant power elements within the generator and load macro-models [8]-[10]. However, the introduced nonlinearities that are defined by the conjugate operator in the complex domain, which is non-analytic, prevent the application of derivative-based numerical algorithms to solve the resulting nonlinear complex circuit. Therefore, in order to allow the application of nonlinear iterative algorithms, such as Newton Raphson (NR), to solve for the operating point of the nonlinear circuit, its complex governing equations are split into their real and imaginary parts. This corresponds to splitting the complex equivalent circuit into its real and imaginary sub-circuits, coupled by controlled sources that can then be linearized and iteratively solved. Most importantly, any power system device can be translated to the circuit domain [11], and further hierarchically combined to build the equivalent circuit of an entire power grid. The derivations of the most prominent power system models can be found in [8]-[13].

The circuit representation of the electric power system provides the opportunity to integrate the resulting equations into any energy management function that includes the power flow equations as constraints such as in state estimation. We have recently demonstrated that both PMU and RTU measurement data can be modeled by equivalent circuits whose parameters are limited by the bounds obtained from interval analysis [16], and thus incorporated within the power grid equivalent circuit without loss of generality.

Assuming a power grid that is fully observable by completely accurate PMUs and following the circuit substitution theorem [14], a PMU can be trivially handled by replacing the measurement device with either a voltage or a current independent source. For instance, if a voltage source model is used, then the current through the voltage source has to be exactly equal to the one measured by the respective PMU. However, incomplete PMU penetration and measurement uncertainties cause discrepancies between the current and voltage in the circuit and the measurements. Therefore, in order to include the non-ideality of current and voltage measurements, we add the conductance ($G_{PMU}$) in parallel to the PMU measurement current source to capture the discrepancy. Minimizing the mismatch current flowing through the added PMU conductances corresponds to minimizing the measurement discrepancy and leads to the optimization problem formulation as given in [16] and used later in this paper. Consequently, the equivalent split-circuit of a PMU device with $N$ terminals is presented in Fig. 1.

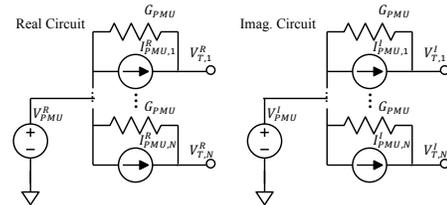

Figure 1. Split-circuit model of a PMU measurement device.

In contrast to PMUs, an RTU measures the magnitudes of the voltage and current ($V_{RTU}$ and $I_{RTU}$) signals as well as the phase angle ($\varphi_{RTU}$) between them. It was shown in [16] that the RTU can be modeled as an injection in terms of bounded admittance state variables, where conductance $G_{RTU}$ supplies or absorbs the real power, while $B_{RTU}$ represents a capacitive or inductive susceptance that adjusts the reactive power. Hence, the governing circuit equations of an RTU device that map the equivalent circuit from Fig. 2 are given as:

$$I_R = G_{RTU}V_R + B_{RTU}V_I \qquad (1)$$
$$I_I = G_{RTU}V_I - B_{RTU}V_R \qquad (2)$$

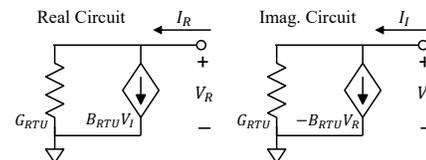

Figure 2. Nonlinear split-circuit of an RTU measurement device.

*B. Formulating the State Estimation optimization problem*

Considering the PMU and RTU measurement data incorporated within the equivalent circuit representation, the optimization problem is formulated to estimate the state of a power system by minimizing the non-idealities in the circuit, namely the currents through the conductances $G_{PMU}$ as explained earlier and deviations of the conductances $G_{RTU}$ and susceptance $B_{RTU}$ from their measured values [16]:

$$\min_{X} \mathcal{F}_e(X) = \left\|I^C_{G_{PMU}}\right\|^2_2 + \left\|G_{diff}\right\|^2_2 + \left\|B_{diff}\right\|^2_2 \quad (3a)$$

subject to split-circuit equations and additional bounds:

$$I_c(X) \equiv \begin{cases} I^R_M + Y_G V_R - Y_B V_I = 0 \\ I^I_M + Y_G V_I + Y_B V_R \end{cases} \quad (3b)$$

$$\underline{X} \leqslant X \leqslant \overline{X} \equiv I_b(X) \leqslant 0 \quad (3c)$$

Where $C \in \{R, I\}$ denotes the real and imaginary components of split-circuit equations, while $Y_G$ and $Y_B$ represent real and imaginary terms of the bus admittance matrix. Current, $I^C_M$ corresponds to the measurement models given for the $i^{th}$ bus as:

$$\begin{bmatrix} I^{R,i}_M \\ I^{I,i}_M \end{bmatrix} = \begin{cases} G^r_{RTU} V^i_R + B^r_{RTU} V^i_I \\ G^r_{RTU} V^i_I - B^r_{RTU} V^i_R & \forall r \in N_{RTU} \text{ if RTU} \\ -I^{R,p}_{PMU} - I^{R,p}_{G_{PMU}} \\ -I^{I,p}_{PMU} - I^{I,p}_{G_{PMU}} & \forall p \in N_{PMU} \text{ if PMU} \end{cases} \quad (4)$$

And $X$ represents a vector of split-circuit state variables:

$$X = [V_C, G_{RTU}, B_{RTU}, V^C_{PMU}, I^C_{PMU}]^T \quad (5)$$

which is bounded by its lower and higher limits ($\underline{X}$ and $\overline{X}$). These limits correspond to setting bounds on the difference between measured and actual/estimated values. Lastly, the non-idealities that are expressed in terms of the measurement mismatch currents for PMUs and distance to the mean for RTU values ($G_m$ and $B_m$) can be written as:

$$I^C_{G_{PMU}} = G_{PMU} \odot (V^C_{PMU} - V^C_T) \quad (6)$$
$$G_{diff} = G_{RTU} - G_m \quad (7)$$
$$B_{diff} = B_{RTU} - B_m \quad (8)$$

It is important to note that in addition to the conventional definition of the state estimation problem, any physics-based and semi-empirical models can be incorporated within the optimization framework without loss of generality. For instance, the BIG load model in [20] can be configured to include the measurement confidence intervals if a sequence of measured grid data is provided and it would remain linear within the equivalent circuit framework.

Next, the Lagrangian function for the SE optimization problem from (3a)-(3c) can be defined in terms of primal and dual variables ($X, \lambda$ and $\mu$) as:

$$\mathcal{L}(X, \lambda, \mu) = \mathcal{F}_e(X) + \lambda^T I_c(X) + \mu^T I_b(X) \quad (9)$$

One of the most prominent methods for handling constrained optimization programs and finding their optimal solution is the Primal-Dual Interior Point (PDIP) method [21]. It obtains the necessary KKT optimality conditions by differentiating (9) with respect to the primal and dual variables, and iteratively solves the resulting equations (10)-(11) while approximating the complementary slackness conditions by (12):

$$\nabla^T_X I_c(X)\lambda + \nabla^T_X I_b(X)\mu = -\nabla_X \mathcal{F}_e(X) \quad (10)$$
$$I_c(X) = 0 \quad (11)$$
$$\mu \odot I_b(X) = -\varepsilon \quad (12)$$

where $\nabla_X I_c(X)$ and $\nabla_X I_b$ are Jacobian matrices, while the average complementary slackness violation $\varepsilon$ from (12) approaches a value close to zero when the iterates reach convergence. Additionally, damping heuristics are applied in order to ensure the feasibility of the iterated variables [21].

Lastly, due to the RTU nonlinearities, the solution ($X^*$) to (10)-(12) is said to be an optimal solution if it further satisfies the second order sufficient condition [21] given by:

$$\tau^T [\nabla^2_{XX} I_c(X^*)]\tau > 0 \quad \forall (\tau \neq 0) \in T_{X^*} \quad (13)$$

where $T_{X^*}$ represents the tangent linear sub-space at $X^*$. Most importantly, the transmission network is defined by the linear constraints and the RTU nonlinearities are introduced locally to each bus. Hence, the second order sensitivity matrix $\nabla^2_{XX} I_c(X^*)$ represents a block diagonal matrix, whose eigenvalues can be determined analytically, namely set of eigenvalues corresponding to each block, which significantly reduce computation of the condition in (13) and is further discussed in Section IV.

### III. EQUIVALENT CIRCUIT PROGRAMMING (ECP)

It was recently shown that constrained optimization problems formulated in terms of equivalent circuit constraints, such as the proposed SE problem formulation, exhibit a unique characteristic [17]-[18]; namely, the complete set of optimality conditions represents the governing equations of an equivalent circuit. When generalized, we can consider this as a class of optimization problems that we refer to as ECPs. Thus, instead of applying the generalized optimization methods to solve for the optimal solution of the SE problem, we further utilize the equivalent circuit formalism behind the SE problem to solve it as a circuit simulation problem.

Even though the circuit simulation formulation for state estimation significantly reduces the nonlinearities of the problem, it remains nonlinear in definitions of RTU circuit models which then also appears in the constraint set of the optimality conditions. Importantly, since the first introduction of the SPICE-like circuit simulators [14]-[15], it has been demonstrated that the simulation of large-scale nonlinear problems requires the complete knowledge of the physical characteristics of the nonlinearities to allow for the development of optimal heuristic algorithms. For instance, it would be intractable to use generalized nonlinear solvers to simulate a billion-node integrated circuit with millions of steep nonlinearities, such as diodes and transistors, without utilizing the knowledge of device physics as it is done in SPICE [15]. Hence, the circuit simulation community has developed efficient models and tools to deal with such nonlinearities that are now leveraged to solve the arising nonlinearities in ECP.

In this section, we first discuss the general relationship between the optimality conditions of the ECP problem and the generalized adjoint network theory. We then show that the power grid equivalent circuit that incorporates measurement data that are coupled with its adjoint circuit will exactly represent the necessary KKT conditions of an ECP problem.

Adjoint circuit theory was explored and applied in the early years of circuit simulation research [14]-[15],[19] and has been largely used for noise analysis [19]. We have recently demonstrated in [17]-[18] that the linear adjoint circuit theory can be generalized for nonlinear circuits at a fixed frequency. Moreover, it was shown in [17] that the governing equations of the adjoint circuit exactly represent the dual equations of the optimality conditions, e.g. (10). Herein, we derive the generalized adjoint circuit equations from TT.

Consider a primal time invariant network $S$ and its adjoint (dual) $\tilde{S}$ defined at a fixed frequency, where the $I$, $X$, $\mathfrak{T}$ and $\lambda$ represent the branch current and state variables of the primal and adjoint networks respectively. From Tellegen's Theorem [19], the primal and adjoint branch currents and state variables need to satisfy the following relationship:

$$I^T \lambda - \mathfrak{T}^T X = 0 \quad (14)$$

Next, let the primal circuit equation have a form of the first order model as given by

$$I = \mathcal{J}(X) X \quad (15)$$

By substituting (15) into (14), the TT can be rewritten as:

$$X^T (\mathcal{J}(X)^T \lambda - \mathfrak{T}) = 0 \quad (16)$$

Hence, for Tellegen's Theorem to remain satisfied, the vector of adjoint currents $\mathfrak{T}$ representing the transformation from primal to adjoint circuit must be defined by:

$$\mathfrak{T} = \mathcal{J}(X)^T \lambda \quad (17)$$

As can be seen from (17), the linear sensitivity matrix $\mathcal{J}(X)$ (linear circuit equations) will result in the linear adjoint circuit, while the nonlinearities of the primal circuit introduce nonlinearities within the adjoint domain. Furthermore, since the excitation sources do not affect the adjoint circuit [19], its operating point given by definition (17) is trivial, namely equal to zero. However, as shown in [18], adding a vector of excitation sources ($\psi_G$) to the adjoint circuit equations corresponds to embedding the negative gradient of an objective function to set the operating point of the adjoint circuit, thereby ensuring the optimality of the primal variables. Consequently, the transformation from (17) is rewritten to include the vector of adjoint excitations:

$$\mathcal{J}(X)^T \lambda = \mathfrak{T} + \psi_G \quad (18)$$

The relationship between the primal and adjoint circuit elements is generalized as given in Table I [17]. Note that herein, the primal and adjoint circuit elements from Table I are considered in terms of their split-circuit representation, however this generalization of adjoint theory also holds for any harmonic or time domain analysis.

In addition to the vector of adjoint excitation sources that ensures the optimality of the respective primal variables, we have shown in [18] that the vector of adjoint currents $\mathfrak{T}$ further enables the control of primal variables. This is done by coupling the adjoint circuit to its control part, modeled in terms of diode circuits as shown in Fig. 3. As it can be seen, the diodes only start conducting if the voltage-controlled voltage sources approach the threshold values set by the variable upper and lower bounds (constraint becomes active). Most importantly, if we approximate the exponential diode models with the hyperbolic functions, its governing equations exactly correspond to the complementary slackness conditions in (12).

Finally, to relate the primal and dual equivalent circuits to the optimality conditions of the ECP problem, consider the governing equations of the primal and adjoint circuits from (15) and (18), and the optimality conditions given by (10)-(12).

The constraints of the optimization problem represent the governing circuit equations, hence the primal problem in (11) corresponds to the governing equations of the primal circuit from (15). Furthermore, the adjoint circuit governing equations represent the dual problem form (10) whereas the vector of adjoint current sources $\mathfrak{T}$ provides the control, while the vector of adjoint excitations ensures optimality.

TABLE I. RELATING THE CIRCUIT ELEMENTS TO ADJOINT (DUAL) DOMAIN

| Primal circuit | | Adjoint circuit |
|---|---|---|
| Independent current source | → | open |
| Independent voltage source | → | short |
| Capacitor | ↔ | Inductor |
| Conductance | → | Conductance |
| Constant Real Power Load | → | Constant Real Power Load |
| Constant Reactive Power Element (Inductive) | ↔ | Constant Reactive Power Element (Capacitive) |
| Objective function gradient | → | Adjoint input source |

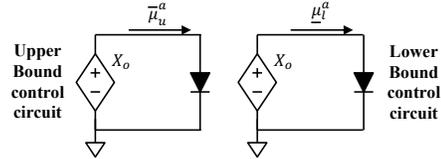

Figure 3. Generalized ECP diode control circuit.

With the relationship between the primal and adjoint circuits fully established, we can apply the derived transformations from Table I to derive the power system equivalent circuit models. For instance, the transmission line π-model is translated to the adjoint domain as shown in Fig. 4. The resulting ECP formulation represents the generic power system optimization framework, whereas depending on the objective of the optimization, only local changes to the circuit models have to be made. Most importantly, the operating point of the derived equivalent circuit exactly represents an optimal solution of the optimization problem that can be obtained using the advanced circuit simulation algorithms [13],[14]-[15]. Techniques such as voltage [12] and admittance limiting [18], as well as diode heuristics [14]-[15] can be used to ensure global convergence and scalability to any-size power systems.

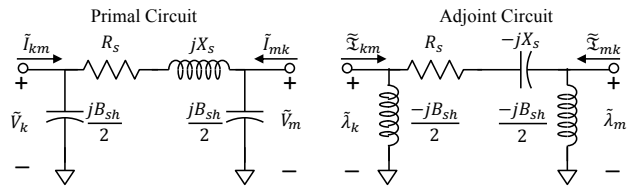

Figure 4. Complex primal and adjoint circuits of a π transmission line model.

## IV. FORMULATING THE STATE ESTIMATION PROBLEM AS AN EQUIVALENT CIRCUIT PROGRAM.

To incorporate the measurement data within the equivalent circuit of an ECP problem, we derive the adjoint split-circuit models of PMU and RTU measurement data that further ensure the minimization of the objective function given in (3a) with respect to the measurement bounds.

We start deriving the adjoint RTU model by finding the sensitivity $\mathcal{J}_{RTU}(X)$ matrix of circuit equations from (1)-(2).

$$\mathcal{J}_{RTU}(X) = \begin{bmatrix} G_{RTU} & B_{RTU} & V_R & V_I \\ -B_{RTU} & G_{RTU} & V_I & -V_R \end{bmatrix} \quad (19)$$

To ensure optimality and bound the RTU admittance state variables, we substitute $\mathcal{J}_{RTU}(X)$ into the generalized definition of adjoint network (18), which further results in the set of governing adjoint RTU circuit equations given as:

$$J_{RTU}(X)^T \lambda + \begin{bmatrix} 0 & 0 & 0 & 0 \\ 0 & 0 & 0 & 0 \\ 1 & -1 & 0 & 0 \\ 0 & 0 & 1 & -1 \end{bmatrix} \begin{bmatrix} \bar{\mu}_G \\ \underline{\mu}_G \\ \bar{\mu}_B \\ \underline{\mu}_B \end{bmatrix} = -2 \begin{bmatrix} 0 \\ 0 \\ G_{diff} \\ B_{diff} \end{bmatrix} \quad (20)$$

As it can be seen from (20), the first two equations represent the adjoint RTU admittance. Furthermore, by setting the adjoint branch currents ($\mathfrak{I}_R$ and $\mathfrak{I}_I$) to be the output currents of the adjoint RTU admittance, we can further write its governing adjoint equations as:

$$\mathfrak{I}_R = G_{RTU}\lambda_R - B_{RTU}\lambda_I \quad (21)$$
$$\mathfrak{I}_I = G_{RTU}\lambda_I + B_{RTU}\lambda_R \quad (22)$$

Importantly, the use of primal and adjoint branch currents to define a model, such as currents from (1)-(2) and (21)-(22), is the typical practice in equivalent circuit modeling. *The respective currents are not the variables of the formulation, but rather an aggregation of the remainder of the system.*

The last two equations represent the constraints added for unknown RTU admittance state variables that further ensure its optimality and control. Furthermore, each of the RTU admittance variables is controlled by the upper and lower bound control circuits, as given in Fig. 3. Lastly, since the RTU model introduces the nonlinearities within the primal and adjoint circuit, it is further linearized by means of the first order Taylor expansion that corresponds to the linearization of the KKT optimality conditions.

Lastly, it should be noted that the sensitivities of the (20) defines a diagonal block of the second order sensitivity matrix $\nabla^2_{XX}I_c(X^*)$ from (13). Hence, it can be shown that tuples of eigenvalues of $\nabla^2_{XX}I_c(X^*)$ that correspond to an RTU bus can be analytically determined to be:

$$2 \times \{\lambda_R^* \pm j\lambda_I^*\} \quad (23)$$

To derive the adjoint split-circuit model of a PMU device from the *equivalent circuit perspective*, consider its primal circuit shown in Fig. 1. First, by applying the established relationships between the primal and adjoint circuit domains from Table I, the PMU voltage sources are shorted, while the current sources are replaced by an open circuit. Next, to ensure the optimality of the current that models the measurement nonidealities, we add the excitation sources to the nodes related to the PMU currents. Lastly, if the measurements are not exact, the PMU voltages and currents are bounded by connecting the additional controlled current sources (see Fig. 5) that couple the adjoint PMU model with the ECP control circuits from Fig. 3. It is important to note that the governing equations that result in the adjoint PMU model from Fig. 5 can be exactly obtained from the optimality conditions (10)-(12).

We can simplify the PMU adjoint circuit by using the circuit perspective to the problem. As shown in Fig. 5, the adjoint currents that control the PMU voltages are shorted. Hence, they do not affect the ECP circuit, and are removed together with the respective complementary slackness conditions without loss of accuracy. Then, the voltage limiting technique [12] is applied to ensure the control of the PMU voltage, while by substitution theorem [14], the current flowing to the ground has to correspond to the removed adjoint currents.

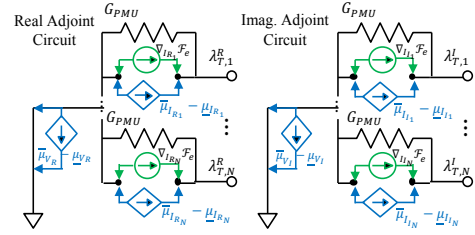

Figure 5. Adjoint split-circuit of a PMU measurement device.

## V. SIMULATION RESULTS

The efficiency and robustness of the proposed ECP framework for power system state estimation are demonstrated by examining several large-scale test cases. This includes the ARPA-E test cases of South Carolina and United States grid (Eastern and Western Interconnects together with ERCOT system) [22], the 70,000 buses Eastern Interconnection test case, as well as the French and European transmission networks (RTE and PEGASE test cases) [23]. The data that further describe the examined benchmarks as well as assigned numbers of measurement devices is presented in Table II.

TABLE II. EXAMINED TEST CASE DATA

| Test Case | Bus [#] | PMU [#] | RTU bus [%] |
|---|---|---|---|
| S. Carolina | 500 | 35 | 93 |
| RTE | 6,515 | 651 | 90 |
| PEGASE | 13,659 | 1,230 | 90 |
| East | 70,000 | 7,000 | 90 |
| USA | 82,507 | 8,160 | 90 |

The derived ECP equivalent circuit models for PMU and RTU devices are incorporated within the C++ prototype version of our ECP circuit simulator. Additionally, the MATPOWER input file is extended to include the measurement data in terms of PMU measured currents and voltages as well as admittance bounds of the RTU equivalent circuit model. Lastly, a MATLAB open source version of the proposed ECP formulation for solving the SE problem is available on: **https://github.com/markojereminov/ECP_based_SE.**

In order to obtain realistic synthetic measurement data and further capture the possible measurement deviations of PMUs and RTUs, the power flow solution ($\bar{X}$) is taken as an accurate measurement for which we add the additional noise as follows:

$$\hat{X}_M = \bar{X} + \sigma_{STD}(2\kappa - 1) \quad (24)$$

where $\sigma_{STD}$ represents the vector of respective standard deviations (see Table III.), and $\kappa$ is a normally distributed random number on an open interval [0,1].

TABLE III. MEASUREMENT STANDARD DEVIATIONS

| RTU Measurements | | | PMU Measurements | |
|---|---|---|---|---|
| Current | Voltage | Power Factor | Current | Voltage |
| 0.4% | 0.4% | 0.6% | 0.02% | 0.02% |

Finally, to include the measurement uncertainties of RTU and inaccurate PMU devices [16], the bounds of $\pm 3\sigma_{STD}$ around the measured values $\hat{X}_M$ are obtained and further used within the equivalent circuit models. Moreover, the values of PMU conductance that models its nonideality ($G_{PMU}$) [16], is set based on the difference in the order of magnitude of $\sigma_{STD}$ of PMU and RTU measurements, namely set to 10 p.u.

Next, to study the effect of random noise introduced within the measurement data (24), and further demonstrate the robustness and accuracy of the proposed ECP approach, 50 sets of measurement data are generated for each of the examined test cases, while keeping the assigned PMU/RTU buses fixed. The developed ECP state estimator prototype is run on a MacBook Pro 2.9 GHz Intel Core i7, and the results obtained are examined by calculating two performance indicators; namely the sum of square of deviations between the accurate and estimated measurements (25), and the maximum absolute deviation as given by (26).

$$\boldsymbol{\sigma}_{ss} = (\widehat{\boldsymbol{X}}_{est} - \overline{\boldsymbol{X}})^T(\widehat{\boldsymbol{X}}_{est} - \overline{\boldsymbol{X}}) \quad (25)$$

$$\boldsymbol{\sigma}_{max} = \max|\widehat{\boldsymbol{X}}_{est} - \overline{\boldsymbol{X}}| \quad (26)$$

The simulation results that include the average values for both performance indicators are presented in Fig. 6 as a function of examined power system sizes.

To further analyze the efficiency of the proposed formulation, we show the runtime comparisons and their average for all of the 50 sets of measurement data as a function of examined system size in Fig. 7. The average values of performance indicators and runtimes of the examined test cases are summarized in Table IV.

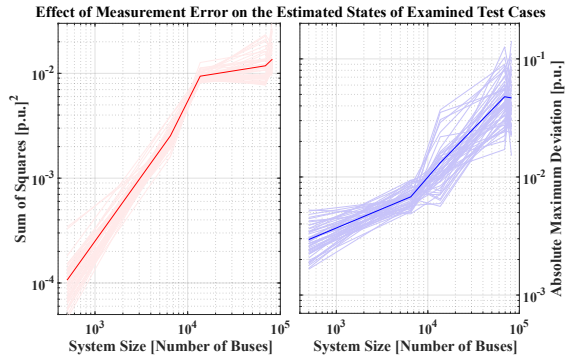

Figure 6. Evaluating the effect of measurement errors to the estimated states.

As can be seen from the presented results in Fig. 6 and Fig. 7 as well as Table IV, the proposed ECP framework successfully and efficiently obtained the power grid state estimates for all of the examined cases. The calculated values of performance factors are sufficiently small, which further indicates that the introduction of accurate PMU measurements within the problem significantly improves the state estimation accuracy. Finally, the efficiency demonstrated by the average runtimes represent the promising improvements that can further lead toward the ultimate goal of real time SE.

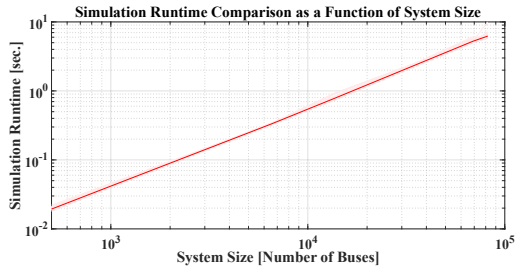

Figure 7. Simulation runtime as a function of examined grid sizes.

TABLE IV. AVERAGE RESULTS

| Test Case | Runtime [sec] | $\sigma_{ss}$[p.u.]$^2$ | $\sigma_{max}$[p.u.] |
|---|---|---|---|
| S. Carolina | 0.0193 | 1.06E-4 | 2.95E-3 |
| RTE | 0.331 | 2.53E-3 | 6.79E-3 |
| PEGASE | 0.780 | 9.39E-3 | 1.31E-2 |
| East | 5.252 | 1.18E-2 | 4.77E-2 |
| USA | 6.236 | 1.36E-2 | 4.67E-2 |

## VI. CONCLUSIONS

In this paper, we presented the ECP formulation for the recently introduced power system state estimation problem defined in terms of equivalent circuit network representation and measurement constraints. It was shown that the power system state can be estimated by solving an ECP circuit simulation problem, without loss of accuracy or generality. Most importantly, the equivalent circuit formalism allows for the understanding and utilizing the physical characteristics of the problem's optimality conditions to develop an efficient, scalable and provably convergent power grid state estimator. Lastly, the introduced framework is generic and can include any physics-based devices models as well as can be applied to distribution systems without loss of generality.

REFERENCES

[1] H. Rudnick, L. Barroso, "Facing uncertainties: the economics of transmission networks", IEEE P&E Magazine, Vol. 14-4, July 2016.
[2] F. C. Schweppe, J. Wildes, "Power system static-state estimation, Part I: exact model", IEEE Trans. on Power Apparatus and Syst., Vol. PAS-89, no. 1, pp. 120–125, Jan 1970.
[3] A. Abur, A. Exposito, "Power System State Estimation: Theory and Implementation", Marcel Dekker, Inc., 2004.
[4] A. Phadke, et.al., "Real time voltage phasor measurements for static state estimation", IEEE Trans. on Power Syst.Vol.104-11, Nov.1985.
[5] M. Zhou, et. al., "An alternative for including phasor measurement in state estimators", IEEE Trans. on Power Syst., Vol.21-4, Nov.2006.
[6] A. S. Costa, A. Albuquerque, et.el., "An estimation fusion method for Including phasor measurements into power system real-time modeling", IEEE Trans. on Power Syst. Vol 28-2, May 2013.
[7] G. Valverde, et. al., "A constrained formulation for hybrid state estimation", IEEE Trans. on Power Syst. Vol.26-3, Aug. 2011.
[8] M. Jereminov, D. M. Bromberg, L. Xin, G. Hug, L. Pileggi, "Improving robustness and modeling generality for power flow analysis," T&D Conference and Exposition, 2016 IEEE PES.
[9] D. Bromberg, M. Jereminov, et. al., "An equivalent circuit formulation of the power flow problem with current and voltage state variables", PowerTech Eindhoven, June 2015.
[10] M. Jereminov, et.al., "An equivalent circuit formulation for three-phase power flow analysis of distribution systems" IEEE T&D, 2016.
[11] A. Pandey, M. Jereminov, et. al., "Unified power system analyses and models using equivalent circuit formulation," IEEE PES Innovative Smart Grid Technologies, Minneapolis, USA, 2016.
[12] A. Pandey, M. Jereminov, G. Hug, L. Pileggi, "Improving power flow robustness via circuit simulation methods," IEEE PES GM, 2017.
[13] A. Pandey, M. Jereminov, M. Wagner, G. Hug, L. Pileggi, "Robust convergence of power flow using Tx-Stepping method with equivalent circuit formulation" XX (PSCC), Dublin, Ireland, 2018.
[14] L. Pileggi, R. Rohrer, C. Visweswariah, Electronic Circuit & System Simulation Methods, McGraw-Hill, Inc., New York, NY, USA, 1995.
[15] W. J. McCalla, "Fundamentals of Computer-Aided Circuit Simulation", Kluwer Academic Publishers, Boston, 1988.
[16] A. Jovicic, M. Jereminov, L. Pileggi, G. Hug, "An equivalent circuit formulation for power system state estimation including PMUs", 2018 North American Power Symposium (NAPS).
[17] M. Jereminov, D. Bromberg, A. Pandey, M. Wagner, L. Pileggi, "Adjoint power flow analysis for evaluating feasibility", IEEE Trans. on Power Syst. (submitted).
[18] M. Jereminov, A. Pandey, L. Pileggi, "Equivalent circuit formulation for solving AC-OPF", IEEE Trans. on Power Systems. (to appear)
[19] S.W. Director, R. Rohrer, "The generalized adjoint network and network sensitivities", IEEE Trans. on Circuit Theory, vol.16, 1969.
[20] M. Jereminov, A. Pandey, H. A. Song, B. Hooi, C. Faloutsos, L. Pileggi "Linear load model for robust power system analysis", IEEE PES ISGT, Torino Italy, September 2017.
[21] S. Boyd, L. Vandenberghe, Convex Optimization, Cambridge University Press, New York, NY, USA, 2004.
[22] A. B. Birchfield, T. Xu, T. Overbye, "Power flow convergence and reactive power planning in creation of large synthetic grids", IEEE Trans. on Power Systems, 2018.
[23] C. Josz, et. al, "AC power flow data in MATPOWER and QCQP format: iTesla RTE snapshots and PEGASE". ArXiv, March 2016.